\RequirePackage[2020-02-02]{latexrelease}
\documentclass[%
reprint,
 amsmath,amssymb,
 aps,superscriptaddress,
]{revtex4-2}
\usepackage{graphicx}
\usepackage{dcolumn}
\usepackage{bm}
\usepackage{xcolor}
\usepackage{hyperref}
\usepackage{multibib}

\begin{document}

\preprint{APS/123-QED}

\title{Direct Search for Dark Matter Axions Excluding ALP Cogenesis in the 63-67 micro-eV Range, with The ORGAN Experiment}
\author{Aaron Quiskamp}
\email{aaron.quiskamp@research.uwa.edu.au}
\affiliation{ARC Centre of Excellence For Engineered Quantum Systems and ARC Centre of Excellence For Dark Matter Particle Physics, Department of Physics, University of Western Australia, 35 Stirling Highway, Crawley WA 6009, Australia.}
\author{Ben T. McAllister}
\email{ben.mcallister@uwa.edu.au}
\affiliation{ARC Centre of Excellence For Engineered Quantum Systems and ARC Centre of Excellence For Dark Matter Particle Physics, Department of Physics, University of Western Australia, 35 Stirling Highway, Crawley WA 6009, Australia.}
\affiliation{ARC Centre of Excellence for Dark Matter Particle Physics,
Swinburne University of Technology, John St, Hawthorn VIC 3122, Australia}
\author{Paul Altin}
\affiliation{ARC Centre of Excellence For Engineered Quantum Systems, The Australian National University, Canberra ACT 2600 Australia}
\author{Eugene N. Ivanov}
\affiliation{ARC Centre of Excellence For Engineered Quantum Systems and ARC Centre of Excellence For Dark Matter Particle Physics, Department of Physics, University of Western Australia, 35 Stirling Highway, Crawley WA 6009, Australia.}
\author{Maxim Goryachev}
\affiliation{ARC Centre of Excellence For Engineered Quantum Systems and ARC Centre of Excellence For Dark Matter Particle Physics, Department of Physics, University of Western Australia, 35 Stirling Highway, Crawley WA 6009, Australia.}
\author{Michael E. Tobar}
\email{michael.tobar@uwa.edu.au}
\affiliation{ARC Centre of Excellence For Engineered Quantum Systems and ARC Centre of Excellence For Dark Matter Particle Physics, Department of Physics, University of Western Australia, 35 Stirling Highway, Crawley WA 6009, Australia.}

\begin{abstract}
The standard model axion seesaw Higgs portal inflation (SMASH) model is a well motivated, self-contained description of particle physics over a range of energy scales that predicts axion dark matter particles to exist within the mass range of $50-200\,\mu$eV. To scan these masses an axion haloscope under a strong constant magnetic field must operate between 12 to 48 GHz. The ORGAN experiment (situated in Perth, Australia) is a microwave cavity axion haloscope that aims to search the majority of the mass range predicted by the SMASH model. Here we present results of Phase 1a, the first experiment to scan and search for axions in the microwave Ku Band. Our initial scan sets a new limit on the coupling of axions to two photons of $g_{a\gamma\gamma}\geq 3\times 10^{-12}\, \textrm{GeV}^{-1}$ over the mass range $63.2$ to $67.1~\mu$eV with $95\%$ confidence. This result is the most sensitive to date in this mass range, sufficient to exclude the well motivated ALP (Axion Like Particle) cogenesis model for dark matter, which adds ALPs to the standard model in the early universe to simultaneously explain the observed baryon and dark matter densities. To attain this level of sensitivity we utilised a TM$_{010}$ cylindrical cavity resonator, scanned between 15.28 to 16.23 GHz through the utilisation of a tuning rod. Measurements were performed over a duration of 3.5 weeks with a 74\% duty cycle, with the resonator coupled to a low noise HEMT amplifier and placed inside a superconducting solenoidal electromagnet of 11.5 Tesla in magnetic field strength. 

\end{abstract}

\maketitle
The nature of dark matter remains one of the major unsolved problems in physics and astronomy with precise cosmological measurements indicating that it accounts for 85\% of all the matter in the Universe \cite{planck,wmap}. In this work we undertake a direct search for one of the prime candidates, the dark matter axion, in a well-motivated mass range \cite{SMASH2017,SMASH2019}.

Axions are hypothetical, massive spin-0 particles that were first postulated as a result of an elegant solution to the strong CP problem in quantum chromodynamics \cite{PQ1977,Weinberg1978,Wilczek1978}. The weakly interacting nature of axions, combined with theoretical predictions and early-Universe production mechanisms before or after inflation \cite{Sikivie1983,Sikivie1983b,Preskill1983,Svrcek_2006,Arvanitaki10,Higaki_2013,Baumann16,Co2020,Co2020b,Co2021,Oikonomou21,Sikivie2021,Sokolov:2021uv,DILUZIO20201,Rodd2021} simultaneously make them a compelling dark matter candidate. To directly detect the dark matter axion we may use a resonant cavity haloscope, which exploits the predicted axion two-photon coupling \cite{Sikivie83haloscope,Sikivie1984}. Under this coupling, axions can convert to photons by interaction with another photon via the inverse Primakoff effect. In a typical haloscope experiment, a strong DC magnetic field is used to create a source of virtual photons for axions to interact with, producing real photons at a frequency corresponding to the rest mass of the axion, $m_a$ \cite{Sikivie83haloscope,Sikivie1984}, with some contribution from the axion velocity. In a typical haloscope, a resonant cavity with a suitable mode geometry is employed to capture these axion-converted photons, and the axion signal is read out with a low noise amplification chain.

The received signal power in such an experiment, due to axion-photon conversion on resonance with the cavity, in the limit $Q_L \ll Q_a$ can be expressed as,
\begin{equation}
    P_{\mathrm{signal}}=\bigg( g_{a \gamma \gamma}^{2} \frac{\rho_{a}}{m_{a}}\bigg)\bigg(\frac{\beta}{1+\beta} B_{0}^{2} V C Q_{L}\bigg). 
    \label{signal_power}
\end{equation}
Here, the axion two-photon coupling strength $g_{a \gamma \gamma}$, local dark matter density $\rho_a \approx 0.45 \, \mathrm{GeV/cm^3}$ (assumed to be all axions) and $m_a$ are parameters set by nature. However, parameters controllable by the experimentalist include the magnetic field strength $B_0$, cavity volume $V$, mode-dependent form factor $C$ and loaded quality factor $Q_L = Q_0/(1+\beta)$, where $Q_0$ and $\beta$ denote the unloaded quality factor and receiver coupling strength respectively. Another controllable parameter is the system noise temperature ($T_S$) of the receiver, which represents the random Nyquist noise in the system in terms of the equivalent effective temperature. To minimise these fluctuations, and to enable use of a superconducting solenoidal magnet, a dilution refrigerator is typically used. Furthermore, since the axion mass and its coupling to photons are \textit{a priori} unknown, experiments must tune (or scan) the haloscope resonant frequency, $\nu_c$ to match the axion frequency, $\nu_a \approx m_ac^2/h$. 
The searchable value of the axion mass in theory may span many orders of magnitude, however current models and astrophysical constraints suggest that the axion mass lies in the $\mu\textrm{eV}-\textrm{meV}$ region, for example the SMASH model predicts $50\leq m_a \leq 200\, \mu \textrm{eV}$\cite{SMASH2017,SMASH2019}. In addition recent QCD lattice simulations also favour $40\leq m_a \leq 180\, \mu \textrm{eV}$, with indications that the mass is close to $65\pm 6\,\mu eV$ ($14.2-17.2$ GHz) \cite{Buschmann2022}, of which the mass range scanned in this experiment is a subset. However, since again the axion mass is unknown, experiments are required to span the largest possible axion mass range, to maximize the prospects for discovery. As such, the relevant figure of merit for axion haloscopes is the allowable rate of frequency scanning \cite{Sikivie2021}, given by \cite{Kim2020}
\begin{equation}
\frac{d f}{d t}=\frac{g_{a \gamma \gamma}^{4}}{\mathrm{SNR}^{2}} \frac{\rho_{a}^{2}}{m_{a}^{2}} \frac{B_{0}^{4} V^{2} C^{2}}{k_{B}^{2} T_{\mathrm{S}}^{2}} \frac{\beta^{2}}{(1+\beta)^{2}} \frac{Q_{L}Q_{a}^2}{Q_{L}+Q_{a}}.
    \label{scan_rate}
\end{equation}
Here, $k_B$ is Boltzmann's constant and $Q_a \sim 10^6$ is the expected signal quality factor due to the axion kinetic energy distribution, most commonly cited as a Maxwell-Boltzmann distribution in the case of an isothermal, virialized halo \cite{isothermal_halo_1990,isothermal_halo_2003}. The axion-photon coupling is parameterised by $g_\gamma$, a dimensionless model-dependent number taking a value of $-0.97$ and 0.36 in two benchmark QCD axion models, the Kim-Shifman-Vainshtein-Zakharov (KSVZ) and the Dine-Fisher-Srednicki-Zhitnisky (DFSZ) models respectively \cite{K79,Zhitnitsky:1980tq,DFS81,SVZ80,Dine1983}. So far, only a handful of experiments have reached KSVZ sensitivity \cite{CAPP2021,Lee2020,Backes:2021wd,Alesini2021}, and only one experiment, the Axion Dark Matter eXperiment (ADMX) has reached the weaker DFSZ sensitivity \cite{Braine2020,bartram2021,ADMX2021}. 

Although most axion experiments target the QCD axion model bands, recent theoretical work suggests more general ALP (Axion-Like Particle) models are possible. For example, cogenesis,  which predicts a much stronger axion-photon coupling as a consequence of adding ALPs to the standard model in the early universe to simultaneously explain the observed baryon and dark matter densities \cite{Co2021,Co2020,Co2020b}. Another example is the recent work on photophilic and photophobic axions, which shows that the parameter space for QCD axions may be much wider than conventional KSVZ and DFSZ axion models suggest \cite{Craig:2018tl,Sokolov:2021uv}.

\begin{figure}[t!]
    \centering
    \includegraphics[width=0.65\linewidth]{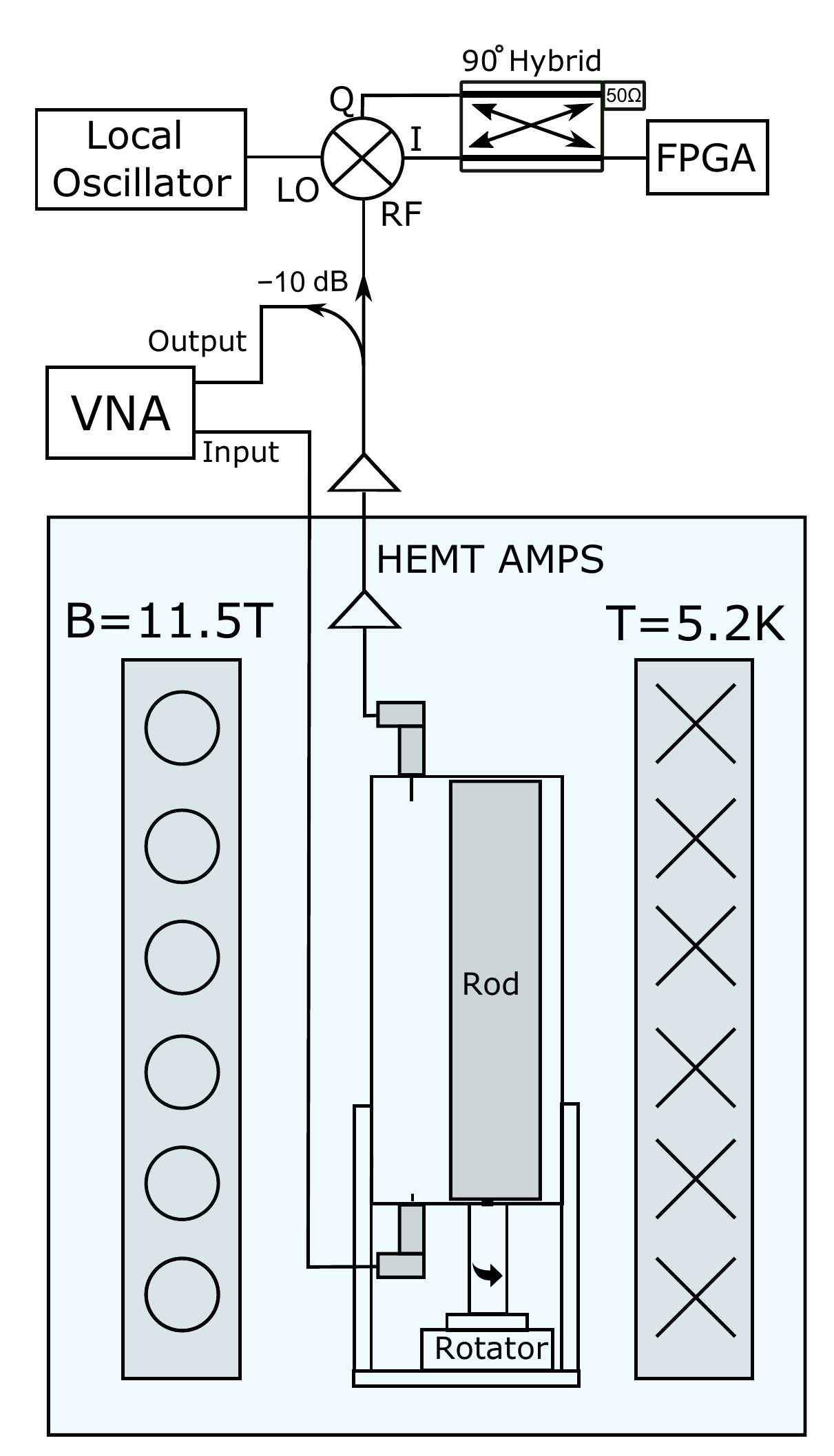}
    \caption{Simple diagram of the Phase 1a experiment. A vector network analyzer (VNA) is used to measure the frequency response of the TM$_{010}$ mode in transmission. An expected axion signal would be coupled out through the HEMT amplifier chain, mixed down using an IQ-mixer and 90 degree hybrid coupler to be sampled by the FPGA digitizer.}
    \label{fig:setup}
\end{figure}

The Oscillating Resonant Group AxioN Experiment (ORGAN) is a microwave cavity haloscope hosted at the University of Western Australia (UWA),  which aims to search for axions in the $62-207\, \mu \textrm{eV}$ ($15-50\, \textrm{GHz}$) mass region. The ORGAN run plan consists of several phases over different regions in the $15-50 \, \textrm{GHz}$ parameter space, with the first 26.5 GHz path-finding run already complete \cite{MCALLISTER201767}. We present the results of Phase 1a (schematic shown in Fig.\ \ref{fig:setup}), which scans for axion masses in the $15.28-16.23$ GHz ($63-67 \,\mu$eV) region of axion-photon coupling parameter space using a tunable $\textrm{TM}_{010}$-based copper conducting-rod resonator. This initial phase of ORGAN serves to test the ALP cogenesis model, operating at a physical temperature $\sim 5.2\textrm{K}$ in a 11.5 T magnetic field, and utilising the best available low-noise High Electron Mobility Transistor (HEMT) amplifiers. Whilst this experiment is capable of placing sensitive limits on axion-photon coupling in its own right, it also serves as a path-finder for future ORGAN phases, which aim to reach the QCD KSVZ and DFSZ model bands.

Compared to lower frequency axion haloscopes, the disadvantage of the cavity resonator approach at the SMASH frequencies of interest is the necessarily small volume, which is of the order of the cube of the axion Compton wavelength, while the scan rate (Eqn.\ \ref{scan_rate}) is proportional to the volume squared.  This presents a significant challenge for haloscopes going forward into the higher mass ranges. We are pursuing various parallel lines of research and development to mitigate this issue, and achieve high sensitivity in the region of interest.

In the supplementary material we detail a path for the cavity resonator technique to be extended to reach KSVZ and DFSZ in the higher mass range.  Subsequent stages of ORGAN will utilise the Phase 1a infrastructure as a test bed for various technologies and techniques required to attain this sensitivity, such as GHz single photon counting \cite{Lamoreaux2013,Kuzmin2018,instruments5030025}, novel cavity designs utilising dielectric boosted sensitivity \cite{Supermode2018,Quiskamp2020}, and utilising multiple cavities simultaneously \cite{XSWisp}. Other experiments which target this mass range, such as MADMAX \cite{MADMAX2017} and ALPHA \cite{ALPHA2019} use other strategies to mitigate this problem, but are both currently in research and development. 


Using Eqn.\ \ref{signal_power} and inserting typical values for our Phase 1a detector, we find $P_{\textrm{signal}}\simeq2.3\times 10^{-25} \, (3.1\times 10^{-26})$ W for a $64\, \mu$eV KSVZ (DFSZ) axion. The total system noise referred to the receiver input is a function of the physical cavity temperature $T_C$, the total added noise from the amplifier chain $T_A$, and the detuning from resonance $\Delta=(\nu-\nu_c)/\Delta\nu_c$, where $\Delta\nu_c$ denotes the cavity linewidth. When the cavity and first amplifier are directly coupled with no isolation, $T_S$ can be expressed as \cite{amp_noise_absorb},

\begin{equation}
	T_S=T_C	\frac{4 \beta}{(1+\beta)^{2}+4\Delta^{2}}+T_A \frac{1+4\Delta^{2}}{(1+	\beta)^{2}+4\Delta^{2}}.
	\label{Tsys}
\end{equation}

For our parameters, the total system noise temperature is orders of magnitude greater in power than that from axion conversion at DFSZ sensitivity.  However, the power generated by a cogenesis ALP is many orders of magnitude greater than the KSVZ or DFSZ axion,  we calculate $P_{\textrm{signal}}\simeq 8.5\times 10^{-21}$ W at $64\,\mu$eV, which is of similar order to the system noise in the ORGAN-Phase 1a experiment. To improve our ability to resolve any signal above these fluctuations, we must average many measurements to improve the signal-to-noise ratio (SNR). For a total integration time $\tau$, the maximum achievable SNR is given by the Dicke radiometer equation: \eqref{SNR}, where $\Delta\nu_a$ is the expected axion linewidth. 

\begin{equation}
    \textrm{SNR}=\frac{P_{\textrm{signal}}}{k_B T_S} \sqrt{\frac{\tau}{\Delta \nu_{a}}}.
    \label{SNR}
\end{equation}

ORGAN-Phase 1a was run for $\sim2.5$ weeks in September 2021 and a further week in January 2022, where typical parameters for the TM$_{010}$ mode include, $Q_L\simeq 3 500$, $C\simeq 0.4$ and $\beta$ between 0.2 and 3. This initial phase was designed as the simplest implementation of the experiment with equipment on hand, the setup of which is shown in Fig.\ \ref{fig:setup}. The cavity is coupled directly to the amplifier to minimize losses, and instead of measuring the readout antenna coupling \textit{in situ} during the experiment, we opted to set the coupling statically at room temperature, and characterise it at as a function of mode frequency, at cryogenic temperatures on separate, dedicated runs directly before and after the data-taking run. The average deviation in $\beta$ between the `before data-taking' and 'after data-taking' coupling characterisation runs amounts to $10.1\%$, with a standard deviation of $7.7\%$. This verifies that we have a good understanding of the coupling, without the need to measure it \textit{in situ}. Our experiment consists of a 32 mm inner diameter copper cavity that houses a 16 mm diameter copper tuning rod, which at a cavity height of 80 mm, gives $V \sim 48$ mL. The tuning rod was machined so that 0.2 mm gaps between the lid and the rod were present at either end of the rod, thus ensuring smooth tuning at cryogenic temperatures. 

\begin{figure}[t]
    \centering
    \includegraphics[width=\linewidth]{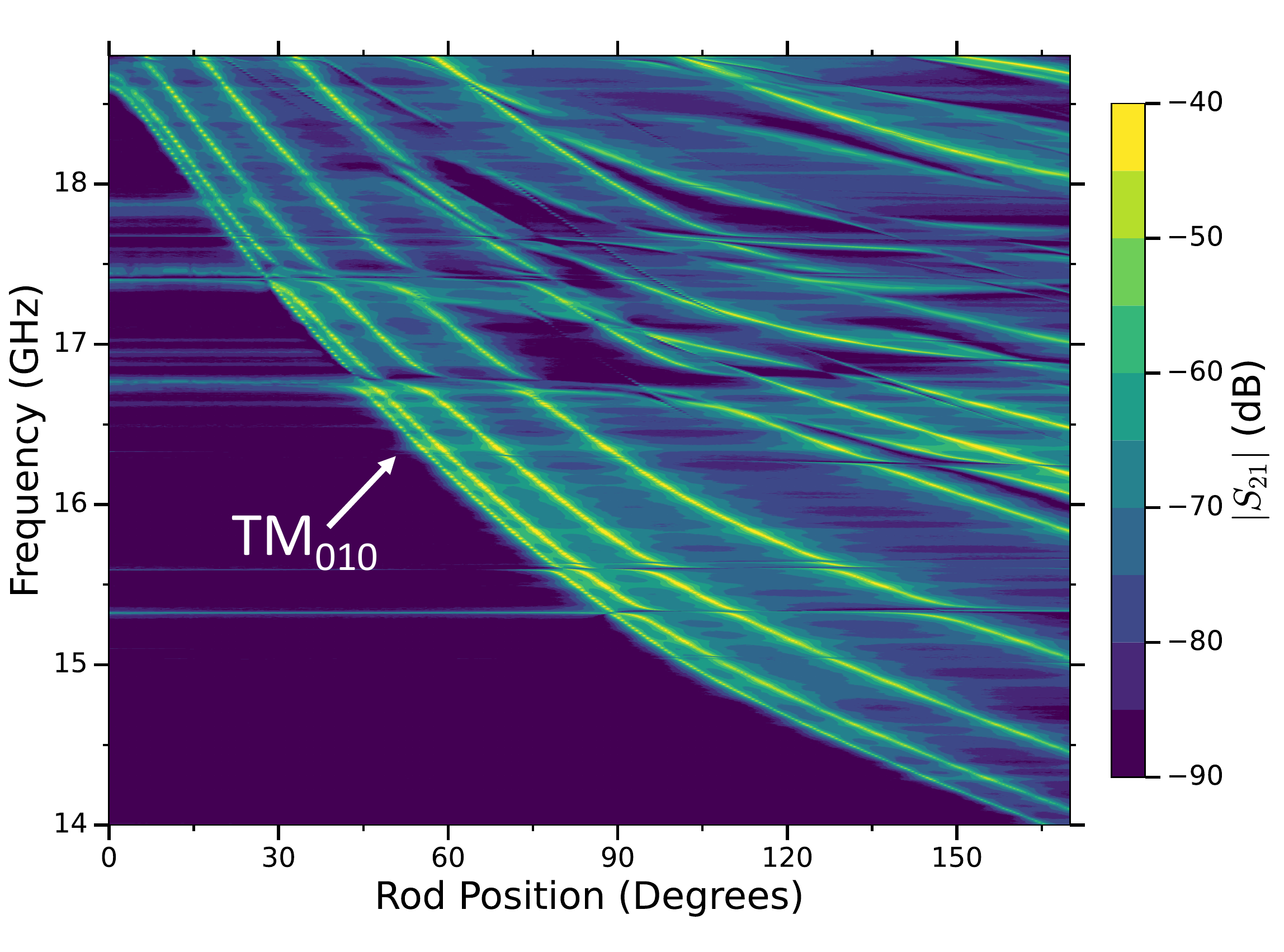}
    \caption{A color-density plot of the transmission coefficient $|S_{21}|$ (dB), as a function of resonant frequency and rod position, with lighter regions representing greater transmission and darker region representing lower transmission. The axion sensitive TM$_{010}$ mode is annotated as the lowest order, tunable mode.}
    \label{fig:mode_map}
\end{figure}

\begin{figure*}
    \centering
    \includegraphics[width=\linewidth]{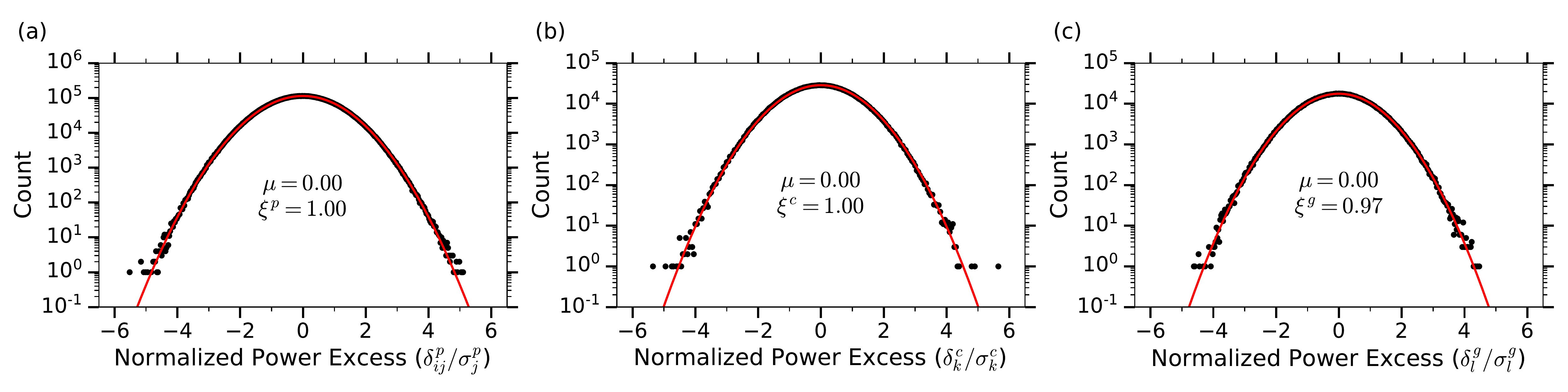}
    \caption{Histograms showing the normalized power spectra at different stages of the analysis procedure, with power excesses shown as black circles and the corresponding Gaussian fits shown in red, where $\mu$ denotes the mean and $\xi$ denotes the standard deviation of the fit. (a) The distribution of the excess power spectra $\delta^p_{ij}$ after being filtered by an SG fit and normalized to $\sigma^p_{j}$, is shown to be perfectly Gaussian. (b) Similarly, the vertical combination of processed spectra preserves the Gaussian nature of the noise, as shown by the distribution of normalized combined spectra $\delta^c_k/\sigma^c_k$. (c) The histogram of all grand spectrum bins   $\delta^g_k/\sigma^g_k$ deviates from the standard normal Gaussian distribution by having a reduced width $\xi^g=0.97$, which comes as a result of the negative SG-filter induced correlations between nearby bins that are co-added. }
    \label{fig:norm_histogram}
\end{figure*}

The coupling was set so that $\beta\simeq2$ was reached at the operating temperature over as much of the target frequency range as possible. Since we had good knowledge of the coupling, with relative changes verified \textit{in situ} by measuring the change in transmission spectrum, the integration time at each cavity step was varied between 45 and 210 minutes depending on the value, so that cogenesis ALPs could be searched for, or excluded over the entire accessible region.

The readout chain, as shown in Fig.\ \ref{fig:setup}, consisted of a vector network analyzer (VNA) that measured (and later utilized) the frequency response of the TM$_{010}$ mode in transmission, a cryogenic pre-amplifer, a directional coupler and a down-conversion stage, which utilised a local oscillator (LO) coupled to an IQ-mixer and 90 degree hybrid coupler. This down-mixing stage achieved image-rejection of the noisy, amplifier-only sideband, giving us the best possible $T_S$. A 250 MS/s digitizer (NI-5761R) was used to sample the output of the hybrid coupler, and the digital data was processed in real time on a field-programmable gate array (FPGA; Xilinx Kintex-7, NI-7935R). A zero-dead-time, hybrid superheterodyne-FFT spectrum analyzer was implemented on the FPGA, which generated a 26,214 point, 12.5-MHz-wide spectrum centered at 45.1 MHz, with bin width $\Delta\nu_b\approx477$ Hz. Frequency tuning was achieved by rotating the off-axis rod with an Attocube ANR240 piezoelectric stepper motor, and steps between successive cavity positions were defined as a fraction of the cavity linewidth. A mode map of angular position versus frequency was made prior to data taking (see Fig.\ \ref{fig:mode_map}), and converted the required frequency step $\sim\Delta\nu_c/5$, to stepper motor steps.

During the total $\sim3.5$ weeks of data taking, 597 resonant frequencies were scanned, amounting to a search window of $\sim700$ MHz between $15.28-16.23$ GHz. Some of the data could not be used due to mode-mixing between the axion-sensitive TM$_{010}$ mode and intruding transverse electric (TE) modes, which arises as a result of the longitudinal symmetry-breaking due to assembly tolerance in the axial position of the tuning rod \cite{Stern2019}. Mode interactions created forbidden frequency ranges between $15.58-15.61$ GHz, $15.65-15.69$ GHz and $15.96-16.15$ GHz, reducing the duty cycle of our search to $\sim74\%$. These regions will be probed in future, more sensitive iterations of the ORGAN experiment. At each cavity position, $Q_L$, $\nu_c$, and $\Delta\nu_c$ were extracted from transmission measurements, so that the LO frequency could be set to $\nu_c - 45.1$ MHz, thus placing the centre of the mode at the centre of the intermediate frequency (IF) window. Although the relatively high IF prevents unwanted spurious noise sources that are common at lower frequencies (IF $\leq$ 10 MHz), there were in fact inescapable, large noise spikes at 40 and 50 MHz due to the device's onboard 10 MHz oscillator, and so we restrict the analysis region to be between $\sim41.2-48.8$ MHz.

We follow the HAYSTAC analysis procedure \cite{Brubaker2017}, which builds upon the pioneering work of \cite{Hagmann_Daw}. The first step is spectrum baseline removal, where the frequency dependent baseline is a combination of the cavity thermal noise and the variable noise and gain of the readout chain. The broadband baseline of each spectrum can be removed using a Savitsky-Golay (SG) filter, which is a digital low-pass filter. The parameters of the SG filter must be chosen so that the passband is flat over large spectral scales, comparable to the baseline, whilst also maximising stop band attenuation over smaller spectral scales, comparable to the axion signal width. As outlined by \cite{Brubaker2017}, the imperfect stop band attenuation of the SG filter induces small negative correlations between neighbouring bins of the same spectrum, and attenuates signals on small spectral scales, like an axion. This effect can be simulated and thus accounted for, but it requires knowledge of the full covariance matrix (see supplementary material). 

\begin{figure*}
    \centering
    \includegraphics[width=0.8\linewidth]{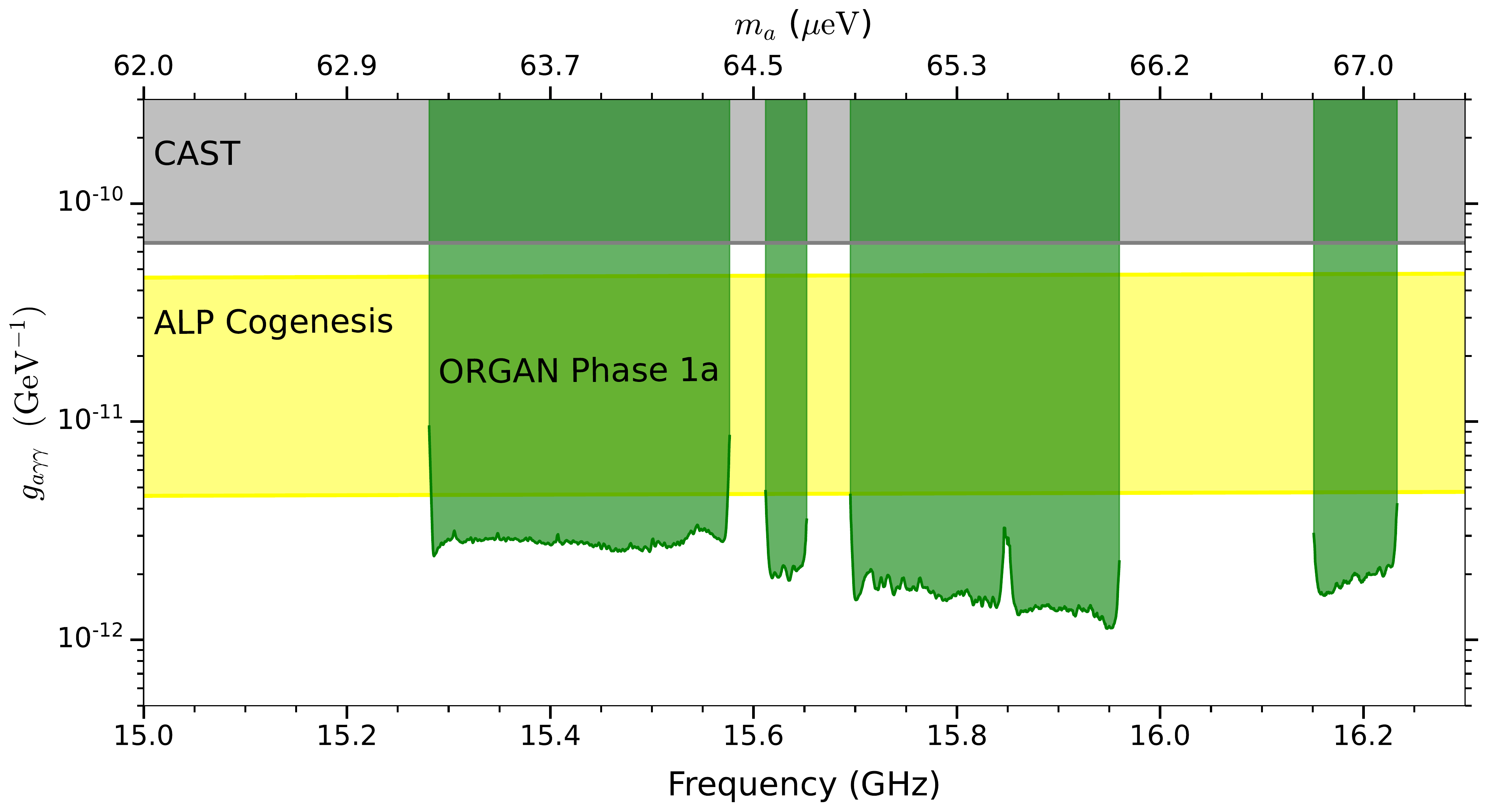}
    \caption{Our $95\%$ confidence exclusion limits on the axion mass-coupling parameter space are shown in green, surpassing the limits set by CAST (grey) \cite{CAST2017} and beyond the ALP-cogenesis model band (yellow) \cite{Co2021,Co2020,Co2020b}). The gaps in the exclusion plot correspond to mode-mixing regions where no axion-sensitive data could be taken. See the supplementary material for a detailed analysis of the fractional uncertainty on these limits, as well as future projections.}
    \label{fig:limits}  
\end{figure*}

Once filtered, the spectra are rescaled according to their axion sensitivity and then vertically combined using a maximum likelihood (ML) weighted sum of contributing spectra to maximize the SNR. As noted, the expected axion signal is thought to have $Q_a\sim10^6$, hence for ORGAN, a 15.6 GHz axion would have a linewidth $\Delta\nu_a\sim 15.6 \, \textrm{kHz}\sim 32\Delta\nu_b$. Therefore we must horizontally combine bins from the vertically combined spectra so that the resulting ``grand spectrum" bin width $\Delta\nu_g$ is $\sim\Delta\nu_a$. Additionally, the expected Maxwell-Boltzmann axion lineshape allows us to further enhance our sensitivity to potential axion signals by optimally filtering sets of 32 consecutive contributing bins according to this lineshape. 


The excess power in the \textit{k}th grand bin is denoted by $\delta^g_k$, and in the absence of correlations the resulting normalized distribution $\delta^g_k/\sigma^g_k$ is expected to have a standard normal distribution. However, as shown in Fig.\ \ref{fig:norm_histogram}, the width of the grand spectrum distribution is reduced to $\xi^g=0.97$, which is a direct result of the negative correlations induced by the imperfect stop-band attention of the SG filter. These negative correlations are accounted for through simulation (details given in the supplementary material), which allows us to place axion exclusion limits in a statistical manner, avoiding Monte Carlo simulations.

We place bin-by-bin exclusion limits on axion-photon coupling by assuming a constant target SNR and re-scaling the sensitivity of each grand bin accordingly \cite{Brubaker2017}. To avoid rescans (which have been shown to have a negligible impact on sensitivity \cite{Brubaker2017})
, we set the candidate threshold at the maximum normalized power excess, $4.6\sigma$, which corresponds to a target SNR of 6.245$\sigma$ at 95$\%$ confidence. As shown in Fig.\ \ref{fig:limits}, we have surpassed the limits set by CAST by over an order of magnitude and excluded ALP-cogenesis over most of the frequency range between $15.28-16.23$ GHz ($63-67\,\mu$eV), with the assumption that axions make up the total local dark matter density. 

ORGAN is the first haloscope experiment to probe axions beyond 10.4 GHz ($43\,\mu$eV) \cite{QUAX_43uev}, into the Ku microwave band, and we have set the most sensitive limits to date on axion-photon coupling in this high-frequency region despite the unfavourable sensitivity scaling. We have detailed the operations and design of our first tunable cavity, with future plans to explore the entire $15-50$ GHz parameter space at QCD model band coupling, with further details given in the supplementary material. Development of quantum readout circuits required to reach this goal, such as GHz single photon counters, is ongoing.

\section*{Acknowledgements}
This work was funded by the Australian Research Council Centre of Excellence for Engineered Quantum Systems, CE170100009 and  Centre of Excellence for Dark Matter Particle Physics, CE200100008, and the Forrest Research Foundation.

\onecolumngrid
\clearpage
\begin{center}
\textbf{\large Supplementary Information for \\
``Direct Search for Dark Matter Axions Excluding ALP Cogenesis in the 63-67 micro-eV Range, with The ORGAN Experiment"}
\end{center}

\section{ORGAN Run Plan}

\begin{figure}[t]
\includegraphics[width=\textwidth]{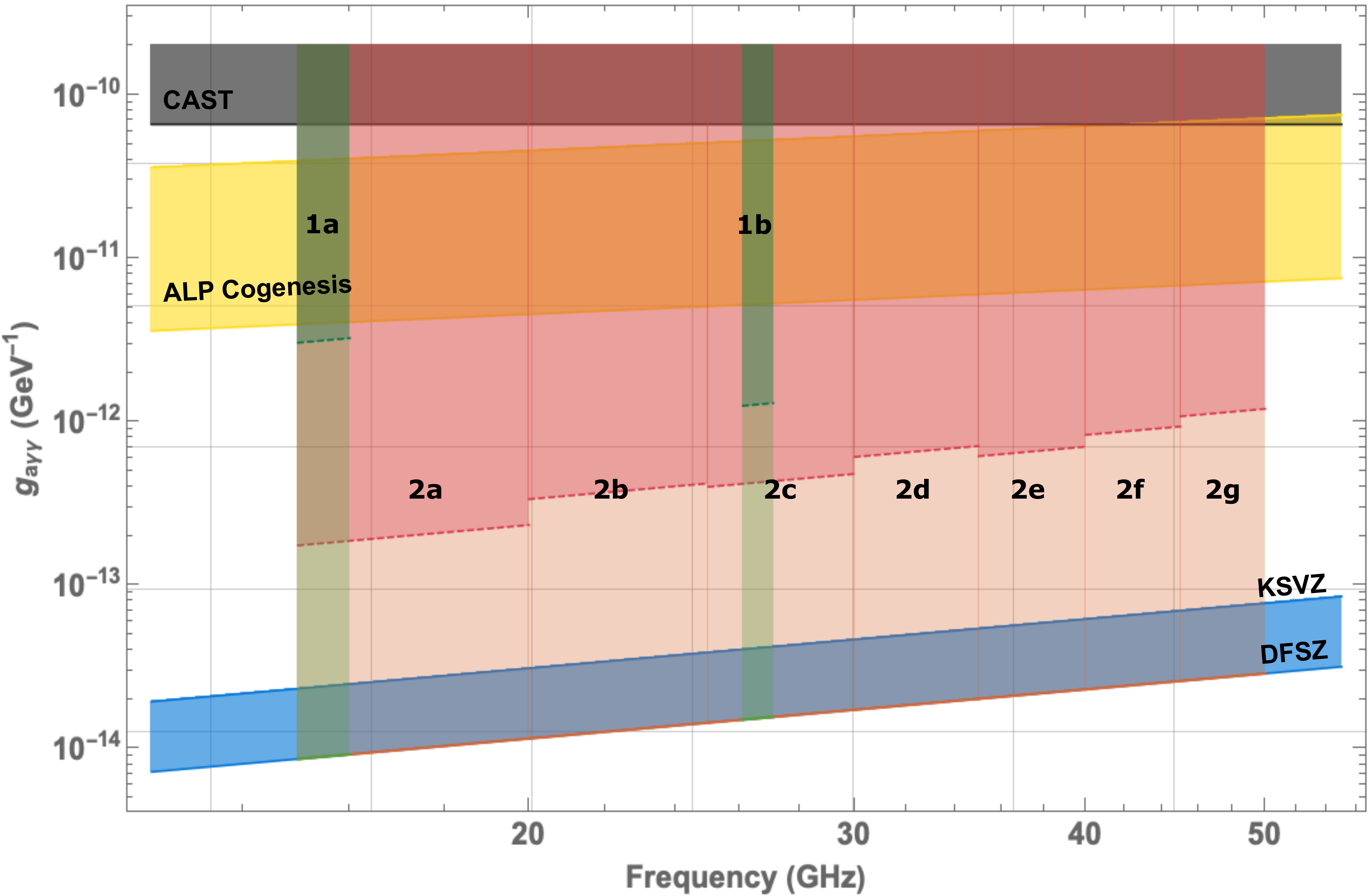}
\caption{Projected exclusion limits based on different assumed parameters. For Phase 1, the darker green limits represent what has already,  been achieved or can be achieved using HEMT-based amplifiers, and resonators, with a form factor of 0.4, and a loaded Q of 10,000.  As discussed, the current plan is to utilize a novel dielectric resonator for 1b - this may change pending research and development outcomes. The lighter green limits represent what can be achieved with single photon counters,  should they be implemented in Phase 1.  Phase 2 assumes a moderate scale up,  utilizing dielectric resonators with a form factor of 0.4,  a loaded quality factor of 30,000,  and 50\% greater volume than the Phase 1 resonators at the same frequency.  The darker red limits represent what can be achieved with quantum limited linear amplifiers, whereas the lighter red limits represent what can be achieved with novel single photon counters with efficiencies on the order of 0.5, and dark count rates on the order of 1000 seconds/photon.  The higher frequency limits assume combination of multiple ($2-4$) cavities. Also shown are the common axion model bands \cite{K79,Zhitnitsky:1980tq,DFS81,SVZ80,Dine1983,Co2021,Co2020,Co2020b}, and the exclusions from CAST \cite{CAST2017}.}
\label{fig:future}
\end{figure}

The run detailed in the main body of the paper constitutes the ORGAN Phase 1a experiment.  On the longer term, the ORGAN run plan consists of two discrete phases,  as outlined in Figure~\ref{fig:future}.

Broadly speaking,  Phase 1 consists of targeted searches utilizing readily available equipment and infrastructure,  such as HEMT-based amplifiers,  and tuning rod based resonators.  Phase 1 searches will also serve as a testbed for various techniques and technologies to be implemented in future ORGAN searches.  Phase 2 consists of more cutting edge searches,  utilizing quantum technology,  multiple cavity synchronization,  novel resonators, and advanced readout techniques to enhance sensitivity.  This is a broad characterization,  as some of these features may be implemented from Phase 1b,  pending research and development.

Phase 1,  comprising Phases 1a and 1b,  commenced in 2021 with Phase 1a.  Phase 1b will operate on some of the same principles as Phase 1a (HEMT amplification,  single cavity),  but with key differences.  Particularly,  we intend to run Phase 1b at mK temperatures,  rather than 4 K,  to implement adjustable resonator coupling,  and to utilize a novel dielectric resonant structure to enhance sensitivity \cite{Quiskamp2020,Supermode2018}.  Phase 1b will cover $26.1-27.1$ GHz,  to test the range of axion masses proposed by the Beck result \cite{BECK2013,BECK2015},  and is projected to commence in 2022,  running for a few months.

During Phase 1,  we will continue research and development for future scale-ups,  including the synchronization of multiple cavities - perhaps utilizing cross-correlation techniques \cite{mcallister2018cross},  superconducting coatings and 3D printed structures \cite{ybco_sc_CAPP},  further novel resonant design,  and single photon counters (SPCs) in the GHz frequency range \cite{SPC_kuzmin_2018,SPC_kuzmin_2021}.  Some or all of these techniques will be implemented in Phase 2,  pending the outcomes of the research and development. 

If efficient GHz SPCs are developed within the timeline of Phase 1,  it is possible that quick re-scans of the 1a and 1b regions will be undertaken with these devices,  to develop expertise in their integration within a haloscope before Phase 2,  and to extend the sensitivity of limits in the region.  Efficient GHz SPCs are seen as the key research and development goal for future ORGAN phases,  as they will allow access to the QCD model bands,  and are well known to be far superior to even quantum limited linear amplification in these regimes \cite{Lamoreaux2013}.

Phase 2 is planned to commence in 2023,  and is broken down into 5 GHz sub-phases,  each lasting $\sim 6-9$ months.  The stages will begin at 15 GHz,  and end at 50 GHz,  scanning the range proposed by the SMASH model \cite{SMASH2017,SMASH2019}.  Phase 2 will employ single photon counters if available by the beginning of the run,  or quantum limited linear amplifiers as a fall-back.  

Research and development for Phase 2 stages will be undertaken on a rolling basis,  where Phase 2a cavities and detectors will be developed during Phase 1,  and later Phase 2 stages will be developed during earlier Phase 2 stages. 

Phase 2 will likely employ dielectric or other novel resonator designs,  and consist of multiple cavities combined at the higher frequency end.  If,  as flagged above,  single photon counters are implemented in Phase 1, this will save scan time in Phase 2,  as 2 GHz of the range will have already been covered at similar sensitivity. 

As with Phase 1,  the use of efficient single photon counters will afford access to DFSZ sensitivity over the entire range,  within the planned timeline,  allowing time for expected maintenance and unplanned stops.  If more traditional quantum-limited linear amplification is employed,  a lower level of sensitivity will be attained,  and efforts will be made to employ sub-quantum limited amplification,  for example via a squeezed state receiver.

For all future projections,  a SNR of 5,  axion density of 0.45 GeV/cc,  magnetic field of 12 T, and physical cavity temperature of $\sim$40-50 mK are assumed,  along with form factors,  volumes and quality factors from models of novel cavities \cite{Supermode2018,Quiskamp2020},  and assumptions of either quantum limited amplification,  or photon counters with efficiencies on the order of 0.5 and dark count rates of 1000 seconds/photon.

As shown in Figure~\ref{fig:future},  ORGAN has the potential to scan the entire $15-50$ GHz range with high sensitivity,  within the decade.

\section{Data Analysis}
The goal of the analysis is to search for significant power excesses above the noise that are consistent with an axion signal. If no such signals are found, limits can placed based on the sensitivity of our detector. We follow the HAYSTAC analysis procedure, with some minor modifications \cite{Brubaker2017b}.

We start by discarding all data files that were not suitable for analysis, for example resonant frequencies that had poor reflection measurement fits, resulting in poor estimates for antenna coupling were discarded. We also remove all single bin contamination present in the intermediate frequency (IF) spectrum, since after very long integration times, the noise floor of the digitzer starts to become visible, with many small spikes scattered throughout the spectrum. The noise floor of the digitizer is measured by terminating the input and averaging for several hours. The bins that contained notable peaks above baseline ($\sigma\geq5$) in the resulting IF spectrum are removed from the real data and replaced with a random value based on the standard deviation and mean in the 40 surrounding bins. In total, 38 out of 15,967 bins in the cropped analysis band were removed and replaced by a random value, thus preserving the Gaussian noise statistics.

\begin{figure}[t!]
    \centering
    \includegraphics[width=\linewidth]{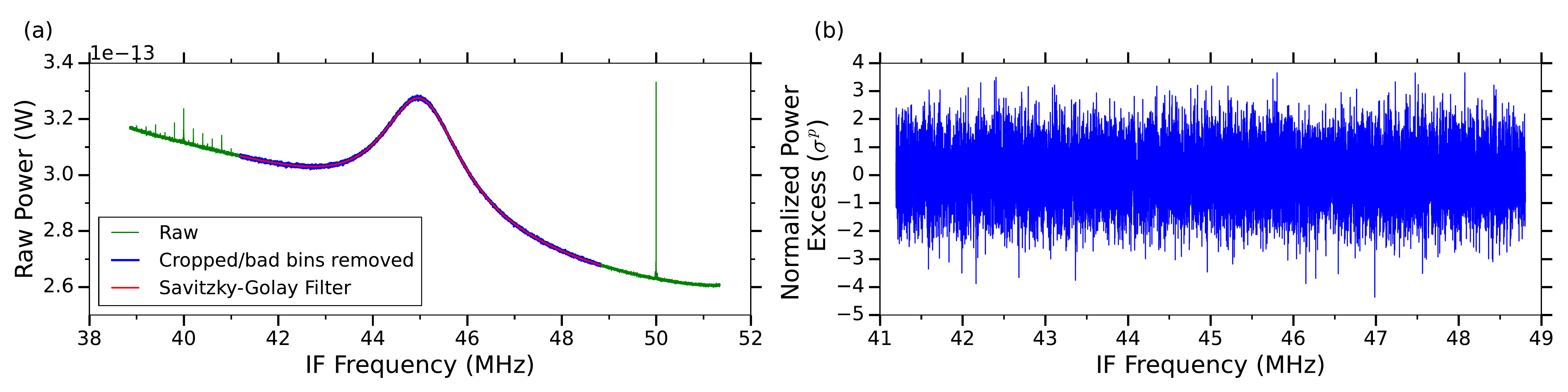}
    \caption{An example of the baseline removal procedure for a single cavity position. (a) The raw spectrum (green) is cropped to the analysis region between $\sim41.2-48.8$ MHz and all narrowband IF contamination is removed (blue). The resulting spectrum is then SG filtered (red). (b) We divide the cropped spectrum by the SG fit and subtracting 1 to give the processed spectrum, which is shown normalized to $\sigma^p$ (blue). }			\label{fig:removal}
\end{figure}

The frequency dependent baseline is removed by a Savitsky-Golay (SG) filter which is parametrised by an impulse window length of $W=450$ bins and polynomial order $d=3$. To obtain a set of dimensionless, normalized excess power spectra, we subtract 1 and in the absence of axion conversion power each bin may be regarded as a Gaussian random variable with mean $\mu=0$ and standard deviation $\sigma^p=1/\sqrt{\Delta\nu_{b}\tau}$, where $\tau$ represents the total integration time and $\Delta\nu_{b}$ is the bin width. The baseline removal procedure is shown in Figure~\ref{fig:removal}, where the raw data has been cropped to the usable bandwidth, corrected for small IF spikes, and subsequently filtered using an SG fit. The now flat excess power is normalized to $\sigma^p$ from which we obtain Figure~\ref{fig:removal}b. 

The excess power in the \textit{j}th IF bin of the \textit{i}th processed spectrum is denoted by $\delta^p_{ij}$, which when normalized to the standard deviation of the corresponding processed spectrum $\sigma^p$, is shown to obey Gaussian statistics, with standard deviation $\xi^p=1$ and $\mu=0$. This result demonstrates that the baseline removal procedure preserves the Gaussian nature of the thermal noise that we are measuring. As the cavity steps through frequency space, there will be a collection of different IF bins from multiple scans that will contribute to the same radio frequency (RF) bin. We follow the general procedure for vertical combination of multiple bins from different scans, using maximum likelihood (ML) weights \cite{Brubaker2017b,Hagmann_Daw}. However, before combination, each bin in each spectrum is re-scaled according to their axion sensitivity. We denote the scaled excess power in the \textit{j}th IF bin of the \textit{i}th spectrum by $\delta_{ij}^s$, which is defined as

\begin{equation}
    \delta_{ij}^s = \frac{\delta_{ij}^p k_B T^S_{ij}\Delta\nu_b}{P^{\textrm{signal}}_{ij}}.
\end{equation}

Here $T^S_{ij}$ is the total system noise temperature referred to the input of the pre-amplifier for a given bin and $P^{\textrm{signal}}_{ij}$ is the expected axion signal power in that bin. The standard deviation is weighted in the same way to obtain a set of scaled, IF bin dependent standard deviations denoted by $\sigma^s_{ij}$ \cite{Brubaker2017b}. The rescaled spectra are then vertically combined using a ML weighted sum of contributing spectra to maximize the SNR \cite{Brubaker2017b,Hagmann_Daw}. The resulting combined excess power in each RF bin is denoted by $\delta^c_k$, where \textit{k} represents the RF bin number. The combined standard deviation $\sigma^c_k$, for the \textit{k}th bin is the quadrature weighted sum of contributing rescaled standard deviations. We expect the distribution of vertically combined bins, normalized to the expected power excess in each combined RF bin to be Gaussian, since independent scans are not correlated. This is shown to be the case, with  a standard deviation of $\xi^c=1$ and mean $\mu=0$. 

We enhance our sensitivity to axion detection by optimally filtering for the expected signal shape. In general, most haloscope searches assume a fully virialized local dark matter velocity distribution, with local circular velocity $v_0 = 220$ km/s. In the galactic frame, the spherically symmetric isothermal model obeys a Maxwell-Boltzmann distribution, and the resulting spectral shape is given by \cite{Brubaker2017b}

\begin{equation}
f(\nu)=\frac{2}{\sqrt{\pi}} \sqrt{\nu-\nu_{a}}\left(\frac{3}{\nu_{a}\left\langle\beta^{2}\right\rangle}\right)^{3 / 2} e^{-\frac{3\left(\nu-v_{a}\right\rangle}{v_{a}\left(\rho^{2}\right)}},
\end{equation}

where $\nu$ and $\nu_a$ denote the measured and axion rest mass frequencies respectively and $\left\langle\beta^2\right\rangle = 3v_0^2/(2c^2)$, where $c$ is the speed of light. The grand spectrum is constructed by horizontal combination of sets of 32 overlapping bin segments, which are ML-weighted sums that take into account the axion lineshape and the standard deviations $\sigma^c_k$, of contributing spectra. The standard deviation for each grand bin is the quadrature weighted sum of contributing standard deviations, which we denote as $\sigma^g_k$. As discussed in the main text, the width of the normalized distribution $\delta^g_k/\sigma^g_k$ is reduced to $\xi^g=0.97$, as a result of the negative correlations induced by the imperfect stop-band attention of the SG filter. The degree of negative correlations between neighbouring bins is completely dependent on the choice of $d$ and $W$, since these parameters decide the cutoff for the passband of the SG filter transfer function. We simulate the effect of the SG filter by comparing SG filtered and non-filtered Gaussian data. We compute the corrected standard deviation, which we denote by $\hat{\sigma}^g_k$, for a weighted sum of correlated Gaussian random variables \cite{Brubaker2017b}. In the case of negative filter-induced correlations, we would expect the sum of off diagonal elements of the covaraince matrix to be negative and hence $\hat{\sigma}^g_k$ will be reduced when compared to the un-corrected $\sigma^g_k$. We expect the ratio of $\hat{\sigma}^g_k/\sigma^g_k$ to be equal to the factor that the normalised grand spectrum width is reduced by, $\xi^g$. Through simulation we find this to be true, with $\hat{\sigma}^g_k/\sigma^g_k = \xi^g = 0.97$ 

\begin{figure}[t]
    \centering
    \includegraphics[width=0.8\linewidth]{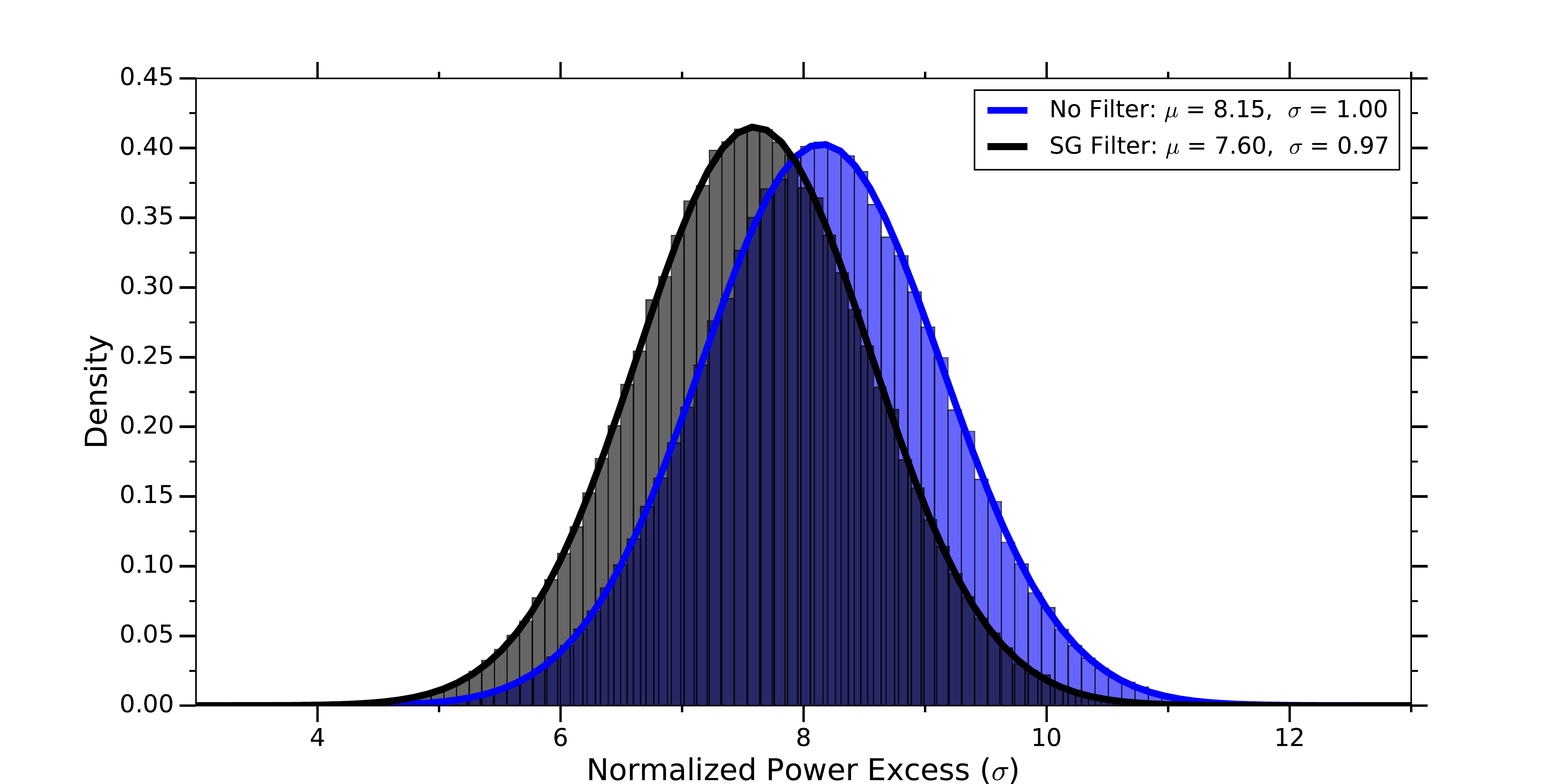}
    \caption{Histograms of the SNR for a synthetic axion signal injected into Gaussian white noise, where one dataset has been SG-filtered (black) and the other has not (blue). After $10^5$ iterations of the simulation discussed in the text, we observe a standard normal distribution for the non-filtered data with $\sigma=1$ and mean SNR $\mu=8.15$, whereas the SG-filtered data results in a Gaussian distribution with reduced width $\sigma=0.97$ and reduced mean SNR, $\mu=7.60$. The negative correlations induced by the SG filter reduced the standard deviation by the same amount as that of the real data, with $\xi^g=0.97$, further demonstrating our understanding of the origin of these correlations. The SG-filter induced attenuation of an axion signal can be expressed as the ratio of the two means $\eta\prime=7.60/8.15=0.93$, which is then normalised to $\eta=\eta\prime/\xi^g=0.96$ for reasons discussed in the text.}  
     \label{fig:SNR_atten}
\end{figure}
 
The negative correlations imposed by the baseline removal procedure will also attenuate a potential axion signal. This reduction in SNR is also simulated, whereby a synthetic axion of known power is injected into 2 analysis pipelines that contain Gaussian white noise. The first pipeline multiplies and then removes a randomly generated baseline using an SG filter, thus simulating the real data analysis procedure, whilst the second control pipeline does not include baseline removal. Once the grand spectrum has been constructed, the normalized power excess $\delta^g_k/\sigma^g_k$, of the synthetic axion is recorded. The simulation is repeated over many iterations, and the results are shown as histograms in Figure \ref{fig:SNR_atten}. The reduction in SNR is represented by the ratio of means between the two pipelines, which we find to be $\eta\prime=0.93$. Since the filter-induced negative correlations also reduce the random noise power fluctuations, as evidenced by the reduced width of the grand spectrum distribution, the total attenuation factor is also reduced by this same amount. We denote the total attenuation in SNR by $\eta = \eta\prime/\xi^g=0.96$. The SNR of each grand bin to a given axion signal is then scaled by this amount to reflect the attenuated SNR, after which limits can be placed in a statistical manner, which are based on the sensitivity of our detector. 

\section{Uncertainty Analysis}
In this section we will quantify our best estimate for the fractional uncertainty in axion-photon coupling for the limits presented in the main text. Given a target SNR and confidence level, limits on $g_{a\gamma\gamma}$ will depend on the bin-by-bin grand spectrum SNR's, that are calculated using the expected axion power at the minimum excludable axion-photon coupling and the noise level of our detector at that frequency. Including only the parameters that carry the greatest uncertainty, we can express the  minimum excludable axion-photon coupling $g^{min}_{a\gamma\gamma}$ in the following way,

\begin{equation}
g_{a\gamma\gamma}^{\min } \propto\sqrt{\frac{T_{\mathrm{A}}}{\frac{\beta}{1+\beta}CQ_L}}.
\label{g_min}
\end{equation}

We calculate the total added noise from the amplifier chain referred to the input of the pre-amplifier, $T_\mathrm{A}$, by using the calibrated frequency dependent effective noise temperature and gain provided by the manufacturer, Low Noise Factory. The fractional uncertainty associated with $T_\mathrm{A}$ is taken to be half the maximum range of $T_\mathrm{A}$ over the frequency region of interest. In future searches, we plan to calibrate this value more precisely \textit{in situ}, thus reducing this uncertainty. As discussed in the main text, the antenna coupling $\beta$, was measured before and after the main data taking run. Therefore we estimate the fractional uncertainty as the mean deviation between these two runs, $\delta\beta/\beta\sim10\%$. The loaded quality factor $Q_L$, is taken from the least squares fitting of transmission measurements to a Fano resonance model, which we find to give reliable and robust estimates of $Q_L$ even in the presence of large asymmetry and noise \cite{fano}. Unlike other experiments which take the standard error from the least squares fit as the uncertainty, we opt for a more conservative approach by taking the average difference between the Fano fitted $Q_L$ and the 3dB width estimate of $Q_L$. We model the frequency dependent form factor $C$ for the TM$_{010}$ mode of the Phase 1a cavity using Finite Element Method (FEM) modelling in COMSOL Multiphysics. Previous searches have taken the uncertainty in $C$ to be $\sim3-5\%$ \cite{ADMX2021,bartram2021}, from which we adopt the upper limit. Our estimates for the fractional uncertainties of the parameters in Eqn.\ \ref{g_min} are tabulated below. 

\begin{table}[h]
\begin{tabular}{cc}
\hline
Source  & \,\, Fractional Uncertainty \\ \hline
$\beta$ & \,\, 0.10                   \\
$Q_L$   & \,\, 0.18                   \\
$C$     & \,\, 0.05                   \\
$T_A$   & \,\, 0.08                   \\ \hline

\end{tabular}
\caption{The dominant sources of systematic uncertainty which are discussed in the text.}
\label{uncert_table}
\end{table}

We vary Eqn.\ \ref{g_min} to find the total fractional uncertainty in $g^{min}_{a\gamma\gamma}$, and substitute the values in the Table \ref{uncert_table} to arrive at Eqn.\ \ref{uncert_eqn} as shown below, where $\bar{\beta}$ is the average coupling over the frequencies scanned. We estimate the total uncertainty in $g^{min}_{a\gamma\gamma}$ to be $\approx 10\%$. 

\begin{equation}
\frac{\delta g_{a\gamma\gamma}^{min}}{g_{a\gamma\gamma}^{min}} \approx \sqrt{\left(\frac{1}{2} \frac{\delta T_{\mathrm{A}}}{T_{\mathrm{A}}}\right)^{2}+\left(\frac{1}{2} \frac{\delta Q_L}{Q_L}\right)^{2}+\left(\frac{1}{2} \frac{\delta C}{C}\right)^{2}+\left(\frac{1}{2} \frac{1}{1+\bar{\beta}} \frac{\delta \beta}{\beta}\right)^{2}} \approx 10\%
\label{uncert_eqn}
\end{equation}

\end{document}